\documentclass[12pt]{iopart}

\usepackage{iopams}
\usepackage{graphicx}
\usepackage{color}
\usepackage{xcolor}
\usepackage{subfigure}
\usepackage{physics}
\usepackage{soul}
\usepackage{color}
\usepackage{bm}
\usepackage{array}
\usepackage{multirow}
\usepackage{lipsum}
\usepackage[normalem]{ulem}
\usepackage{verbatim}
\usepackage{nccmath}
\usepackage{algorithm2e}
\RestyleAlgo{ruled}
\usepackage[pdftex,colorlinks=true,
	pdfstartview = FitV,
	linkcolor    = linkcolor,
	citecolor    = linkcolor,
	urlcolor     = linkcolor,	
	hyperindex   = true,
	hyperfigures = false]{hyperref}
	
\definecolor{linkcolor}{rgb}{0,0,0.6}

\usepackage{cleveref}

\begin{document} 

\title{Species interconversion of deformable particles yields transient phase separation}

\author{Yiwei Zhang}
\address{Department of Physics and Materials Science, University of Luxembourg, L-1511 Luxembourg, Luxembourg}
\author{Alessandro Manacorda}
\address{CNR Institute of Complex Systems, Uos Sapienza, Piazzale A. Moro 5, 00185 Rome, Italy}
\address{Department of Physics and Materials Science, University of Luxembourg, L-1511 Luxembourg}
\author{\'Etienne Fodor}
\address{Department of Physics and Materials Science, University of Luxembourg, L-1511 Luxembourg, Luxembourg}

\begin{abstract}
We consider a dense assembly of repulsive particles whose fluctuating sizes are subject to an energetic landscape that defines three species: two distinct states of particles with a finite size, and point particles as an intermediate state between the two previous species. We show that the nonequilibrium synchronization of sizes systematically leads to a homogeneous configuration associated with the survival of a single species. Remarkably, the relaxation towards such a configuration features a transient phase separation. By delineating and analyzing the dominant kinetic factors at play during relaxation, we recapitulate the phase diagram of species survival in terms of the parameters of the size landscape. Finally, we obtain a hydrodynamic mapping to equilibrium by coarse-graining the microscopic dynamics, which leads to predicting the nature of the transitions between various regimes where distinct species survive.
\end{abstract}

\maketitle


\section{Introduction}

The phase separation of binary mixtures is arguably the simplest type of spatial organisation~\cite{Hohenberg1977, Bray1994}. This phenomenology can be found in a wide spectrum of systems ranging from metallic~\cite{Bocquet1996,Dehosson2001} and glassy~\cite{Bernu87pra,Grigera01pre} alloys to human populations~\cite{Schelling1971}. In a biological context, it has recently been reported that some living cells exploit phase separation to form membraneless organelles~\cite{Hyman2014, Shin2017}, showing how nonequilibrium energy fluxes robustly regulate spatial organization at small scales~\cite{Weber19rpp}.

In equilibrium, the phase separation of multicomponent mixtures can be rationalized by analyzing the free energy in terms of some density fields~\cite{Shrinivas2021, Thewes22prl}. Far from equilibrium, despite the absence of any free energy, various de-mixing mechanisms can still be delineated by considering, for instance, (i)~species with different mobilities and/or temperatures~\cite{Grosberg15pre, Weber16prl, Ilker20prr}, (ii)~species interconverting with rates that break the local detailed balance~\cite{Alston2022, berthin2024}, or (iii)~deformable particles with driven sizes corresponding to various species~\cite{Togashi2019, Zhang23prl, manacorda2023, pineros2024, Dzubiella2024}. These mechanisms lead to either a standard phase separation where the demixed phases simply diffuse [e.g., cases (i-ii)], or richer scenarios associated with the propagation of such phases [e.g., waves in case (iii)].

In some systems, the phase separation is not maintained in steady state, but rather corresponds to a transient relaxation. A seminal example is given by the equilibrium Ising model~\cite{Ising1925}: for the Glauber dynamics at low temperature~\cite{Glauber1963}, a disordered configuration first undergoes a phase separation and then reaches a homogeneous steady state with broken symmetry. A similar scenario can be found in chiral systems~\cite{Latinwo2016, Stillinger2023}, where a disordered mixture of enantiomers undergoes a transient phase separation, before reaching a homochiral steady state. For some living systems, it has been argued that intracellular phase separation can be regarded as a transient state, maintained over a long duration by some periodic driving~\cite{yan2022condensate, Charras2022}.

A theoretical description of transient phase separation is already well established for equilibrium dynamics~\cite{Hohenberg1977, Bray1994}. Far from equilibrium, the hydrodynamic theories of multicomponent systems capture distinct relaxation scenarios~\cite{Odor2008}. For example, some theories inspired by surface catalysis~\cite{Ziff1986} consider the effect of species annihilation upon collisions~\cite{Bramson88prl, Zhuo1993, Brown97pre}: here, transient domains eventually relax to a homogeneous configuration, which corresponds to the survival of a single species. For chiral systems, variants of the Cahn-Hilliard model~\cite{cahn1958free, Li20jstatmech} have inspired some hydrodynamic descriptions that capture enantioselective relaxation~\cite{Anisimov2021, Longo23pnas}. For deformable particles, it remains to build hydrodynamic theories that accurately describe the role of transient phase separation in the competition between species.

In this paper, we analyze the phenomenology of an assembly of interconverting deformable particles in two spatial dimensions. To this end, we consider that the fluctuating sizes of repulsive particles evolve in a specific landscape defining three species [Fig.~\ref{Fig:1}(a)]: particles with a finite radius (either type $A$ or type $B$), and point particles (type $\varnothing$). We focus on the regime where the point particles $\varnothing$ are metastable, so that the system essentially behaves like a binary mixture for the particles $(A,B)$. We reveal that the relaxation towards a homogeneous configuration, corresponding to the survival of a single species, entails a transient phase separation. We identify the dominant kinetic factors at play during such a transient and rationalize the corresponding phase diagram of species survival in terms of the landscape parameters.

The paper is organized as follows. After introducing our model [Sec.~\ref{sec:model}], we study the phase diagram in terms of landscape parameters [Sec.~\ref{sec:ps}], and recapitulate these observations with a hydrodynamic theory [Sec.~\ref{sec:hydro}]. Overall, our results show that dense assemblies of deformable particles subject to a size landscape entail a rich phenomenology, and we elucidate how the transient phase separation controls the species survival.


\section{Particle deformation and species populations}\label{sec:model}

We consider the dynamics of $N$ deformable particles in two spatial dimensions. Each particle $i$ is identified by its position $\mathbf{r}_i$ and its internal degree of freedom reaction $\sigma_i$. In all what follows, we refer to $\sigma_i$ as the {\em reaction coordinate}, by analogy to some chemical reactions, defined so that $|\sigma_i|$ effectively determines the particle radius. The overdamped Langevin dynamics of positions reads
\begin{equation}\label{eqn:particle_based_model_r}
    \dot{\mathbf{r}}_i = -\mu_\mathbf{r} \sum_{j \in \partial i}\partial_{\mathbf{r}_i} U(a_{ij})+ \sqrt{2\mu_\mathbf{r} T}{\boldsymbol\xi}_i \ ,
    \quad
    a_{ij} =\frac{\abs{\mathbf{r}_i-\mathbf{r}_j}}{\abs{\sigma_i}+\abs{\sigma_j}} \ ,
\end{equation}
where $\mu_\mathbf{r}$ is the position mobility, and $T$ the temperature of the surrounding thermostat. The vectorial Gaussian white noise ${\boldsymbol\xi}_i$ is isotropic with unit variance and zero mean. The pairwise potential $U$ describes the repulsion between neighboring particles: $U(a) = a^{-12} - 2a^{-6} + 1$ if $a<1$ and $U(a)=0$ otherwise; the sum over $j \in \partial i$ thus refers to the neighbors interacting that satisfy $a_{ij}<1$. The dynamics of ${\bf r}_i$ is coupled to the dynamics of $\sigma_i$ as
\begin{equation}\label{eqn:particle_based_model_sig}
    \dot{\sigma}_i = \sum_{j \in \partial i}\mu_\sigma\Big[\varepsilon(\sigma_j-\sigma_i)-\partial_{\sigma_i}U (a_{ij}) \Big] -\mu_{\sigma}\partial_{\sigma_i}V(\sigma_i)+\sqrt{2\mu_{\sigma}T}\eta_i \ ,
\end{equation}
where $\mu_\sigma$ is the reaction-coordinate mobility. While $\sigma_i \in \mathbb{R}$, we recall that the radii $\vert \sigma_i \vert$ cannot be negative. The Gaussian white noise $\eta_i$ has unit variance and zero mean, and is uncorrelated with $\bm\xi_i$. The harmonic coupling induces an attractive interaction ($\varepsilon>0$), which we refer to as {\em synchronization}, between the coordinates $(\sigma_{i},\sigma_j)$ of neighboring particles. Such a linear synchronization is arguably the simplest type of term to mimic the equivalent of a communication between particles; for instance, see~\cite{Milster23pre,Dzubiella2024} for further details on how such a  mechanism emerges between deformable colloids.

Although the synchronization depends on space (with connectivity changing in time), it has no counterpart in the position dynamics, so that it drives the dynamics out of equilibrium. In the absence of synchronization ($\varepsilon=0$), the system corresponds to an assembly of passive deformable particles~\cite{Brito18prx} or responsive colloids~\cite{Baul21jpcm}, which relax towards Boltzmann statistics with weight $\exp(-\beta U_{\rm tot})$, where $U_{\rm tot} = \sum_{i,j<i}U(a_{ij}) + \sum_i V(\sigma_i)$ and $\beta=1/T$. When two particles overlap, the repulsive potential $U$ leads them to move away and shrink their sizes, respectively because of the gradient terms in Eqs.~\eqref{eqn:particle_based_model_r} and~\eqref{eqn:particle_based_model_sig}.

\begin{figure}
    \centering
    \includegraphics[width=\linewidth]{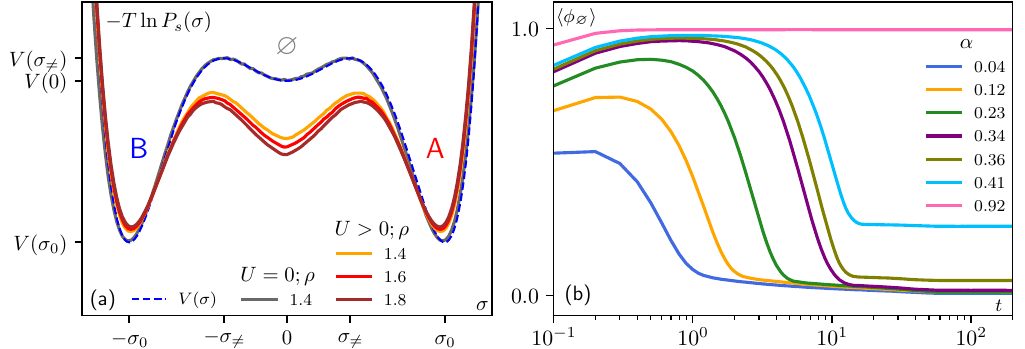}
    \caption{(a)~Log-distribution $- T \ln P_s$ of reaction coordinate $\sigma$ in equilibrium ($\varepsilon=0$). In the absence of repulsion ($U=0$, solid gray line), the distribution follows the Boltzmann weight $e^{-\beta V}$ (dashed blue line) in terms of the one-body potential $V$ [Eq.~\eqref{eqn:one_body_potential}]. In the presence of repulsion ($U>0$), the local minimum corresponding to the metastable state ($\sigma=0$) gets deeper as the density $\rho$ increases. Parameters: $v_0=60$, $\alpha=0.12$, $\mu_\mathbf{r}=1$, $\mu_\sigma=0.1$, $T=5$, $\sigma_0=1$, $\sigma_{\ne} =0.4$ and $L=100$.
    (b)~Time evolution of the averaged population $\langle \phi_\varnothing \rangle$ [Eq.~\eqref{eq:phi}] of point particles $\varnothing$ under different values of the metastability parameter $\alpha$ [Eq.~\eqref{eq:alpha}], as measured over $100$ trajectories. 
    Same parameters as in~(a), with $\rho=1.4$ and $0.04<\alpha<0.92$.}
     \label{Fig:1}
\end{figure}

In addition to the pair repulsion $U$, each coordinate $\sigma_i$ is subject to the one-body potential $V(\sigma_i)$ [Fig.~\ref{Fig:1}(a)]:
\begin{equation}\label{eqn:one_body_potential}
    V(\sigma) = v_0 \bigg(\frac{\sigma}{\sigma_0}\bigg)^2 \bigg[ \bigg(\frac{\sigma}{\sigma_0}\bigg)^4 - \frac 3  2 (1 + \gamma) \bigg(\frac{\sigma}{\sigma_0}\bigg)^2 + 3\gamma \bigg] \ ,
    \quad
    \gamma = \bigg(\frac{\sigma_{\ne}}{\sigma_0}\bigg)^2 \ ,
\end{equation}
where $v_0=1$ is an energy scale. The potential $V$ embodies an effective landscape that constrains the statistics of reaction coordinates $\sigma_i$, and therefore determines the population of various species with distinct $\sigma$. In practice, such a landscape exhibits (i)~two degenerate global minima located at $\pm\sigma_0$, (ii)~two degenerate local maxima at $\pm\sigma_{\ne}$, and (iii)~a metastable local minimum at $\sigma=0$. These features lead to distinguish three species: (i)~two stable species, which we refer to as {\em enantiomers} in analogy to models of chiral particles~\cite{Latinwo2016,Stillinger2023}, denoted by $A$ (for $\sigma > \sigma_{\ne}$) and $B$ (for $\sigma<-\sigma_{\neq}$), and (ii)~a metastable species for $\abs{\sigma}<\sigma_{\ne}$, which we refer to as {\em point particles} and denote by $\varnothing$. We characterize the species populations in terms of
\begin{equation}\label{eq:phi}
    \phi_\varnothing(t) = \int_{-\sigma_{\ne}}^{+\sigma_{\neq}} d\sigma P(\sigma,t) \ ,
    \quad
    \phi_A(t) = \int_{+\sigma_{\neq}}^{+\infty} d\sigma P(\sigma,t) \ ,
    \quad
    \phi_B(t) = \int_{-\infty}^{-\sigma_{\neq}} d\sigma P(\sigma,t) \ ,
\end{equation}
where we have introduced the empirical distribution of reaction coordinates:
\begin{equation}
    P(\sigma,t) = \frac 1 N \sum_{i=1}^N \delta(\sigma-\sigma_i(t)) \ .
\end{equation}
The conservation of particle number enforces $\phi_\varnothing(t) + \phi_A(t) + \phi_B(t) = 1$ at all times. The statistical properties of the system are obtained by simulating the dynamics in Eqs.~\eqref{eqn:particle_based_model_r} and~\eqref{eqn:particle_based_model_sig}. The trajectories evolve with an Euler-Maruyama method~\cite{Gardiner2009} using an adaptive time step. The total duration $t_M$ of a simulation is chosen to reach a steady state, and its typical range is $50 < t_M < 200$ depending on the microscopic parameters.

The metastability of point particles is determined by the relative depth $\alpha$ of the potential wells:
\begin{equation}\label{eq:alpha}
    \alpha=\frac{V(\sigma_{\ne})-V(0)}{V(\sigma_{\ne})-V(\sigma_0)} =
    \frac{\gamma^2(3-\gamma)}{(1-\gamma)^3} \ .
\end{equation}
In the regime $\alpha<1$ (i.e., $\gamma<1/3$), the potential well of enantiomers $(A,B)$ is always deeper than for point particles $\varnothing$. Consequently, in the absence of synchronization ($\varepsilon=0$), the steady-state statistics defined by $P_s(\sigma)=\lim_{t\to\infty}\langle P(\sigma,t)\rangle$, where $\langle\cdot\rangle$ refers to an average over noise realizations, shows a higher population of $(A,B)$ than of $\varnothing$ [Fig.~\ref{Fig:1}(a)]. The repulsive potential $U$ penalizes the overlap between particles, so that it leads to an increase in the population of $\varnothing$ with respect to the case $U=0$. In the presence of synchronization ($\varepsilon>0$), changing the relative potential depth $\alpha$ drastically affects the species populations. After a transient relaxation, $\langle\phi_\varnothing\rangle$ reaches a plateau value that increases with $\alpha$ [Fig.~\ref{Fig:1}(b)]. At high enough $\alpha$, this plateau value saturates to $1$, showing that all particles are point-like. In all what follows, we focus on the regime $\alpha<0.34$, for which $\langle\phi_\varnothing\rangle\ll\langle\phi_A+\phi_B\rangle$ at large times: the steady-state population of enantiomers $(A,B)$ is overwhelmingly dominant compared to that of point particles $\varnothing$. In this regime, the non-monotonic behavior of $\langle\phi_\varnothing(t)\rangle$ entails a rich phenomenology, associated with transient phase separation, as discussed in the next Sections.


\section{Transient phase separation}\label{sec:ps}

In this Section, we examine how the transient phase separation determines the steady-state configuration. We first describe how the topology of phase boundaries affects the relaxation for a symmetric landscape $V(\sigma)$. Then, we discuss how the various kinetic factors regulate the competition between species for an asymmetric landscape $V_\lambda(\sigma)$.

\begin{figure}
    \centering
    \includegraphics[width=\linewidth]{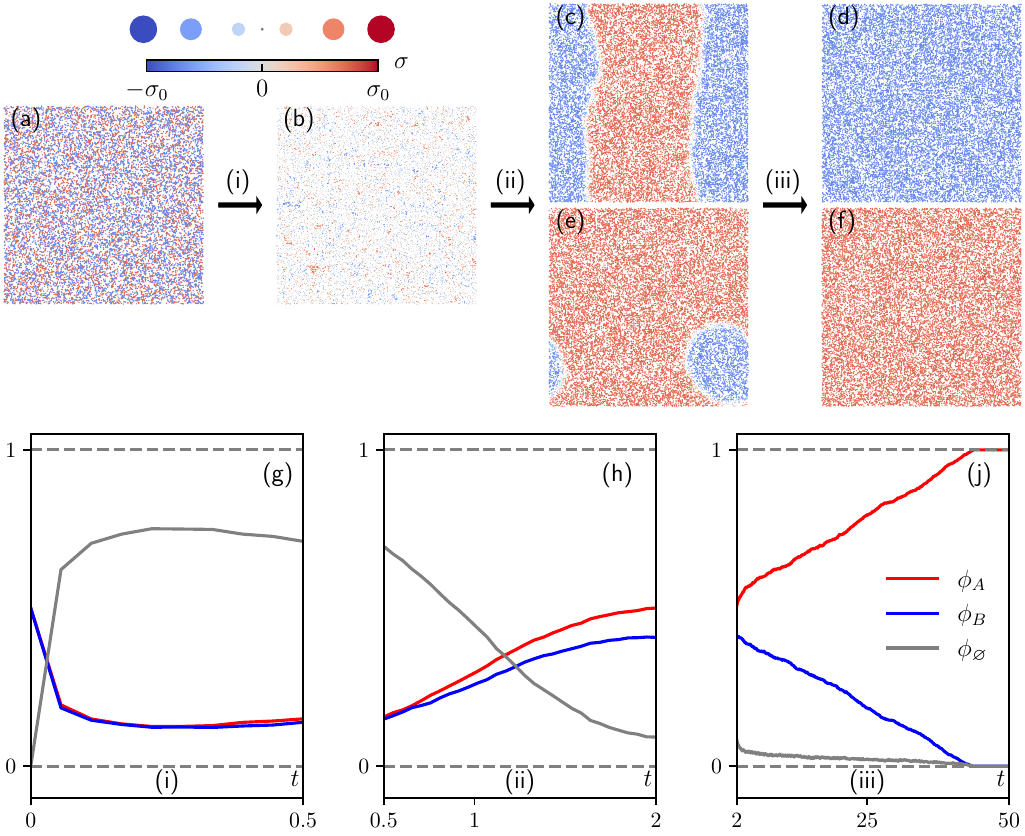}
    \caption{Three-step relaxation: (i)~the particles $(A,B)$ shrink into point particles $\varnothing$ due to synchronization, from (a)~a disordered configuration to (b)~a configuration dominated by particles $\varnothing$; (ii)~particles $\varnothing$ are converted into particles $(A,B)$ forming domains that grow into (c,e)~a phase-separated configuration; (iii) domains of $A$ and $B$ compete to end up with (d,f)~a homogeneous configuration of either $(A,B)$.
    (g,h,j)~The three stages (i-ii-iii) can be clearly identified from the time-evolution of the populations of particles $(A,B,\varnothing)$. Parameters as in Fig.~\ref{Fig:1}, with $\rho=1.4$ and $\alpha=0.12$.
    }
    \label{Fig:2}
\end{figure}

\begin{figure}
    \centering
    \includegraphics[width=.8\linewidth]{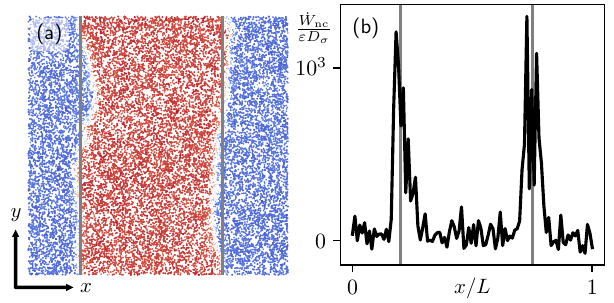}
    \caption{Departure from equilibrium. (a)~Phase-separated configuration of particles $(A,B,\varnothing)$, same color code as in Fig.~\ref{Fig:2}. (b)~The non-conservative work rate $\dot W_{\rm nc}$ [Eq.~\eqref{eq:epr}], shown here as averaged over the $y$ direction, is higher at the boundaries between domains (grey lines) where the synchronization force is predominant. Parameters as in Fig.~\ref{Fig:2}.
    }
    \label{Fig:3}
\end{figure}

\subsection{Symmetric landscape $V(\sigma)$: Phase topology determines the survival of species}

The system always relaxes towards a homogeneous state without any spatial structure. In the absence of synchronization ($\varepsilon=0$), the steady state is disordered with a spatially homogeneous distribution of particles $(A,B,\varnothing)$ [Fig.~\ref{Fig:1}(a)]. In the presence of synchronization ($\varepsilon>0$), the steady state corresponds to a homogeneous profile where almost all particles are in the same state. For a symmetric landscape $V(\sigma)$ [Eq.~\eqref{eqn:one_body_potential}], this state corresponds to enantiomers $A$ or $B$, with equal probabilities.

The simulation is initialized with uniform distribution of positions in the periodic square and $\sigma_i = \pm \sigma_0$ with equal probability. Before reaching the homogeneous state, the relaxation goes through three successive stages [Figs.~\ref{Fig:2}(a-f)]: (i)~initial shrinkage of particles, (ii)~formation and growth of disconnected domains, and (iii)~competition between connected domains. During the initial shrinkage, almost all particles relax towards the state $\varnothing$ [Fig.~\ref{Fig:2}(b)], yielding sharp increase of $\phi_\varnothing$ and decrease of $\phi_A+\phi_B$ [Fig.~\ref{Fig:2}(g)]. Then, during domain growth, the point particles $\varnothing$ are turned into either one of the enantiomers $(A,B)$, leading to a steady decrease of $\phi_\varnothing$ and increase of $\phi_A+\phi_B$ [Fig.~\ref{Fig:2}(h)]. Specifically, separated domains, made of $A$ (resp.~$B$) that have survived the initial shrinkage, locally convert $\varnothing$ into $A$ (resp.~$B$). Once domains have invaded the whole system, the vast majority of point particles $\varnothing$ have become $(A,B)$. Finally, during domain competition, interactions between particles at the domain interfaces lead some domains to coarsen and others to recoil. Eventually, a single domain invades the whole system, yielding only one of the enantiomers $(A,B)$ to survive. In the case where $A$ (resp.~$B$) survives, $\phi_A$ (resp.~$\phi_B$) increases steadily until it reaches $1$, while $\phi_B$ (resp.~$\phi_A$) and $\phi_\varnothing$ decrease and converge towards $0$ [Fig.~\ref{Fig:2}(j)]. Due to fluctuations, the homogeneous domain can potentially switch between $A$ and $B$ (i.e., it does not correspond to an absorbing state), yet we discard such rare events in what follows.

During domain growth and domain competition, the system's behavior is reminiscent of the phase separation observed between liquids~\cite{Hohenberg1977, Bray1994}. Remarkably, the final fate of the system, where only $A$ or $B$ survives, can actually be anticipated from the topology of the phase boundaries. Indeed, domains with positive curvature (e.g., {\em bubbles} of $B$ in a sea of $A$) are quickly eliminated [Figs.~\ref{Fig:2}(e,f)]. In contrast, domains with a band-like structure have a longer lifetime, although they eventually relax towards a homogeneous state [Figs.~\ref{Fig:2}(c,d)]. In practice, we expect that the synchronization between nearby particles plays an essential role in regulating the dynamics of domains. To confirm this effect, we consider the non-conservative work rate $\dot W_{{\rm nc},i}$ produced per particle:
\begin{equation}\label{eq:epr}
    \dot W_{{\rm nc,}i} = F_{\mathrm{nc},i}\circ\dot{\sigma}_i \ ,
\end{equation}    
where $\circ$ refers to a Stratonovich product. Substituting the dynamics of reaction coordinates [Eq.~\eqref{eqn:particle_based_model_sig}] into Eq.~\eqref{eq:epr}, we deduce
\begin{equation}
    \dot W_{{\rm nc,}i} = \mu_{\sigma}F_{\mathrm{nc},i} (F_{\mathrm{nc},i} + F_{\mathrm{c},i}) + \mu_\sigma T \frac{d F_{{\rm nc},i}}{d\sigma_i} \ ,
\end{equation}  
where the conservative and non-conservative forces, respectively $F_{\mathrm{c},i}$ and $F_{\mathrm{nc},i}$, read
\begin{equation}
    F_{\mathrm{nc},i} =\varepsilon\sum_{j \in \partial i}(\sigma_j-\sigma_i) \ ,
    \quad
    F_{\mathrm{c},i} = - \partial_{\sigma_i} \Big[V(\sigma_i) + \sum_{j \in \partial i} U(a_{ij})\Big] \ .
\end{equation}
We recall that the non-conservative nature of $F_{\mathrm{nc},i}$ stems from the lack of any counterpart term in the position dynamics [Eq.~\eqref{eqn:particle_based_model_r}]. For a phase-separated profile, $\dot W_{{\rm nc},i}$ is higher at the boundaries between domains (where nearby particles have different reaction coordinates) than in the bulk (where the coordinate distribution is uniform). This feature confirms the crucial role of synchronization in regulating domain coarsening through their boundaries [Fig.~\ref{Fig:3}]. Remarkably, the major influence of nonequilibrium forces at boundaries is qualitatively identical to the well-established scenario in many-body dynamics of self-propelled particles~\cite{Fodor22arcmp}, despite the different non-conservative forces at play in the two cases.


\begin{figure}
    \centering
    \includegraphics[width=\linewidth]{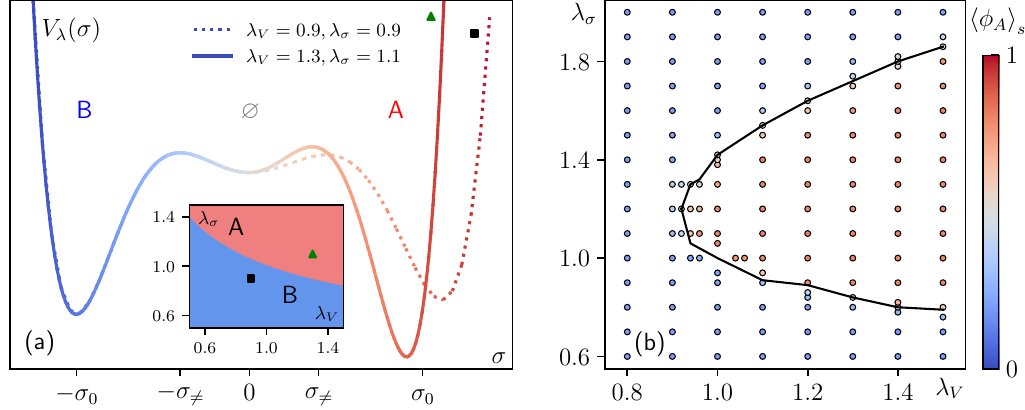}
    \caption{(a)~Asymmetric landscape $V_\lambda$ of reaction coordinate $\sigma$ [Eq.~\eqref{eq:v_lam}] in cases where either particle $A$ (solid line) or particle $B$ (dashed line) is most stable. (Inset)~The relative stability of $(A,B)$, given in terms of the landscape parameters $(\lambda_\sigma, \lambda_V)$, follows from comparing the conversion rates between particles [Eq.~\eqref{eqn:rate_ratio}]. The symbols (green triangle and black square) refer to the parameter values of $V_\lambda$.
    (b)~Phase diagram for the steady-state population $\langle\phi_A\rangle_s$ of particles $A$ in terms of $(\lambda_\sigma,\lambda_V)$. The solid black line corresponds to $\langle\phi_A\rangle_s=0.5$. Parameters as in Fig.~\ref{Fig:2}.
    }
    \label{Fig:4}
\end{figure}

\subsection{Asymmetric landscape $V_\lambda(\sigma)$: Kinetic factors of competition between species}

We now examine the case where the symmetric landscape $V(\sigma)$ [Fig.~\ref{Fig:1}(a)] is replaced by the asymmetric landscape $V_\lambda(\sigma)$ [Fig.~\ref{Fig:4}(a)] defined as
\begin{equation}\label{eq:v_lam}
    V_\lambda(\sigma)=\begin{cases}
            V(\sigma) & {\rm for} \; \sigma<0 \ ,
            \\
            \lambda_V V(\lambda_\sigma\sigma) & {\rm for} \; \sigma>0 \ ,
    \end{cases}
\end{equation}
where ${\bm\lambda}=(\lambda_V,\lambda_\sigma)$ is a set of positive scaling factors, which respectively control the potential well and the typical radius of $A$ particles. Therefore, tuning $\bm\lambda$ is a route towards controlling the relative stability of the enantiomers $(A,B)$. Indeed, the phase diagram of the stationary population $\langle\phi_A\rangle_s = \lim_{t\to\infty}\langle\phi_A(t)\rangle$ distinguishes parameter regimes where either $A$ or $B$ survives [Fig.~\ref{Fig:4}(b)]; note that $\phi_A$ is here defined by replacing $\sigma_{\ne}$ with $\sigma_{\ne}/\lambda_\sigma$ in Eq.~\eqref{eq:phi}. The boundary between these regimes crosses the point ${\bm\lambda}=(1,1)$, where $(A,B)$ have equal survival probabilities, as expected.

Our aim is to identify the dominant kinetic factors favoring either $A$ or $B$, and rationalize accordingly the non-trivial behavior of the phase boundary. We observe that the three-step relaxation, reported for the symmetric landscape $V(\sigma)$ [Fig.~\ref{Fig:2}], carries over to the asymmetric case $V_\lambda(\sigma)$. Let us first assume that the topology of the transient domains controls the steady state as in the symmetric case: bubbles tend to shrink and disappear [Figs.~\ref{Fig:2}(e,f)]. In practice, such a topology results from the formation and growth of disconnected domains [stage (ii) in Fig.~\ref{Fig:2}]: once these domains connect, they adopt a specific topology that shapes their competition [stage (iii) in Fig.~\ref{Fig:2}]. Therefore, we hypothesize that the dominant kinetic factor is the rate of conversion from $\varnothing$ to $(A,B)$ that is at play during domain growth. Specifically, we expect the domains of $A$ (resp.~$B$) grow faster whenever the conversion rate to $A$ (resp.~$B$) is higher, resulting in a higher probability of finding bubbles of $B$ (resp.~$A$) in a sea of $A$ (resp.~$B$), and eventually favors the survival of $A$ (resp.~$B$) only.

To evaluate the conversion rates, we consider a mean-field version of our model [Eq.~\eqref{eqn:particle_based_model_sig}] by (i)~assuming that all particles have the same reaction coordinate $\sigma=\sigma_i$, and (ii)~neglecting the effect of repulsion ($U=0$), yielding
\begin{equation}\label{eq:mf}
    \dot\sigma = - \mu_\sigma \partial_\sigma V_\lambda + \sqrt{2\mu_\sigma T} \eta \ .
\end{equation}
The conversions between species amounts to noise-activated transitions between minima of the landscape $V_\lambda$. Given that Eq.~\eqref{eq:mf} describes an equilibrium dynamics, we deduce that the rates of such transitions follow Kramers' escape formula~\cite{Gardiner2009}. Considering the local minimum $\sigma_{\rm m}$ and maximum $\sigma_{\rm M}$, the corresponding rate $k(\sigma_{\rm m}\to\sigma_{\rm M})$ reads
\begin{equation}
    k(\sigma_{\rm m}\to\sigma_{\rm M}) \sim \sqrt{\left\vert V''(\sigma_{\rm m})V''(\sigma_{\rm M}) \right\vert} e^{-\beta(V_\lambda(\sigma_{\rm M})-V_\lambda(\sigma_{\rm m}))} \ ,
\end{equation}
where $V''=d^2 V_\lambda/d\sigma^2$. The net rates of conversion from $\varnothing$ to either $A$ or $B$ follow as
\begin{equation}
\begin{aligned}
    \omega_{\varnothing A} &= k(0^+\to\sigma_{\neq}/\lambda_\sigma) - k(\sigma_0/\lambda_\sigma\to\sigma_{\neq}/\lambda_\sigma) \ ,
    \\
    \omega_{\varnothing B} &= k(0^-\to-\sigma_{\neq}) - k(-\sigma_0\to-\sigma_{\neq}) \ .
\end{aligned}
\end{equation}
Within our hypothesis, which assumes that the conversion rates determine the system's fate, the steady state is made of either $A$ or $B$ with equal probabilities whenever $\omega_{\varnothing A} = \omega_{\varnothing B}$. This condition yields a specific relation between the scaling parameters:
\begin{equation}\label{eqn:rate_ratio}
    \lambda_\sigma^2 = \frac{e^{\beta(\lambda_V-1)\bar V}}{\lambda_V} \frac{\sqrt{2(1/\gamma-1)}-e^{\beta(1-\alpha)\bar V}}{\sqrt{2(1/\gamma-1)}-e^{\beta\lambda_V(1-\alpha)\bar V}} \ ,
    \quad
    \bar V = (1-\gamma)^3 \ ,
\end{equation}
where $(\alpha,\gamma)$ are defined in Eqs.~\eqref{eqn:one_body_potential} and~\eqref{eq:alpha}. The relation in Eq.~\eqref{eqn:rate_ratio} yields a monotonic boundary line in the space ${\bm\lambda}=(\lambda_\sigma,\lambda_V)$ [inset of Fig.~\ref{Fig:4}(a)]. This boundary crosses the symmetric point ${\bm\lambda}=(1,1)$, and favors $A$ particles at large $\lambda_V$, consistently with the relative stability of $(A,B)$ in the landscape $V_\lambda$ [Fig.~\ref{Fig:4}(a)]. Yet, the boundary obtained from numerical simulations [Fig.~\ref{Fig:4}(c)] of our original model [Eqs.~\eqref{eqn:particle_based_model_r} and~\eqref{eqn:particle_based_model_sig}] features a re-entrance not captured by Eq.~\eqref{eqn:rate_ratio}: $B$ survives both at small and large $\lambda_\sigma$. Therefore, we deduce that some factors other than domain growth determine species survival at large $\lambda_\sigma$.

\begin{figure}
    \centering
    \includegraphics[width=\linewidth]{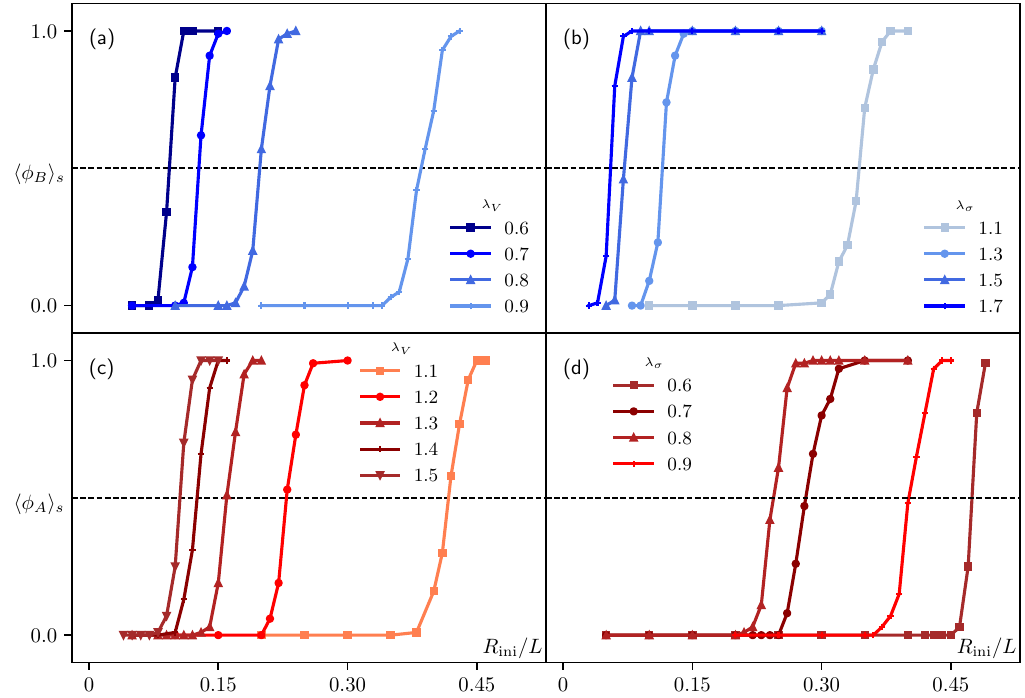}
    \caption{Survival probability $\langle\phi_A\rangle_s$ (resp.~$\langle\phi_B\rangle_s$) as a function of the size of the initial bubble $R_{\rm ini}$ of particles $A$ (resp.~$B$) in a sea of particles $B$ (resp.~$A$). The black dashed line refers to $\langle\phi_A\rangle_s=0.5$ (resp.~$\langle\phi_B\rangle_s=0.5$). Parameters as in Fig.~\ref{Fig:2}, with $\lambda_\sigma=1$ in panels (a,c) and $\lambda_V=1$ in panels (b,d).
    }
    \label{Fig:5}
\end{figure}

During domain competition, bubbles of $A$ (resp.~$B$) can grow in a background of $B$ (resp.~$A$) for some values of ${\bm\lambda}$, in contrast to the symmetric case ${\bm\lambda}=(1,1)$ where bubbles always recoil [Figs.~\ref{Fig:2}(c,d)]. We systematically quantify this effect with an initially phase-separated profile: a single bubble of $A$ (resp.~$B$) is surrounded by $B$ (resp.~$A$). We numerically determine the radius of the bubble over which the survival probability $\langle\phi_A\rangle_s$ (resp.~$\langle\phi_B\rangle_s$) exceeds $1/2$. Either decreasing $\lambda_V<1$ or increasing $\lambda_\sigma>1$ reduces the critical radius of $B$ [Figs.~\ref{Fig:5}(a,b)], while increasing $\lambda_V>1$ reduces the critical radius of $A$ [Fig.~\ref{Fig:5}(c)]: the corresponding regimes of $(\lambda_\sigma, \lambda_V)$ are consistent with the survival of species reported in our phase diagram [Fig.~\ref{Fig:4}(b)]. For $\lambda_\sigma<1$, the behavior of the critical radius of $A$ is non-monotonic [Fig.~\ref{Fig:5}(d)]: this effect potentially stems from the fact that $A$ particles are actually bigger than $B$ ones in this regime.

In short, we have identified two major kinetic factors at play in controlling the relaxation towards a homogeneous configuration: (i)~the conversion of point particles $\varnothing$ into enantiomers $(A,B)$, and (ii)~the expansion of domains with positive curvatures. The relative importance of these factors determines the survival of the species. In practice, our phase diagram [Fig.~\ref{Fig:4}(b)] is consistent with assuming that (i) and (ii) respectively dominate for $\lambda_\sigma<1$ and $\lambda_\sigma>1$.


\section{Coarse-graining from particles to fields} \label{sec:hydro}

In this Section, we propose a derivation of the field theory describing the hydrodynamics of our model by coarse-graining the microscopic dynamics. Such a coarse-graining amounts to an equilibrium mapping, which leads to some analytical predictions for the transitions between various regimes of species survival.

\subsection{Mapping to equilibrium field theory}

To obtain a hydrodynamic description of our model, we start by simplifying the microscopic dynamics [Eqs.~\eqref{eqn:particle_based_model_r} and~\eqref{eqn:particle_based_model_sig}]. First, we neglect the repulsion in the dynamics of positions ${\bf r}_i$ as
\begin{equation}\label{eqn:effective_dyn_r}
    \dot{\mathbf{r}}_i = \sqrt{2\mu_{\mathbf{r}}T}{\boldsymbol\xi_i} \ .
\end{equation}
This simplification amounts to discarding the role of interactions in the position dynamics, which essentially assumes that the dynamics of reaction coordinates $\sigma_i$ alone regulate the collective effects. Second, we assume that particles interact only if they are at the same position:
\begin{equation}\label{eqn:mf_particle_model_2body}
    \sum_{j \in \partial i} \Big[ \varepsilon(\sigma_j-\sigma_i)-\partial_{\sigma_i}U (a_{ij}) \Big] \simeq \sum_{j=1}^N \Big[\varepsilon(\sigma_j-\sigma_i)-\partial_{\sigma_i}U(a_{ij}) \Big]\delta(\mathbf{r}_j-\mathbf{r}_i) \ .
\end{equation}
Third, inspired by the approach in~\cite{Zhang23prl, banerjee2024}, we use the approximation
\begin{equation}\label{eq:approx}
    \sum_{j=1}^N (\partial_{\sigma_i} U)  \delta(\mathbf{r}_j-\mathbf{r}_i) \simeq (\partial_{\varphi}U) \sum_{j=1}^N \frac{d\varphi}{d\sigma_j} \delta(\mathbf{r}_j-\mathbf{r}_i) \ ,
\end{equation}
where we have introduced the packing fraction $\varphi = \pi \sum_i(\sigma_i/L)^2$ in terms of the system size $L$, and assumed that $\partial_{\varphi}U$ is constant. Substituting our approximations [Eqs.~\eqref{eqn:mf_particle_model_2body} and~\eqref{eq:approx}] into the dynamics of reaction coordinates [Eq.~\eqref{eqn:particle_based_model_sig}] yields
\begin{equation}\label{eqn:effective_dyn_sigma}
    \dot{\sigma}_i = -\mu_\sigma\partial_{\sigma_i} \Phi(\sigma_i|\,\overline\sigma, \rho) + \sqrt{2\mu_\sigma T}\eta_i \ ,
\end{equation}
in terms of the effective landscape
\begin{equation}\label{eq:land}
    \Phi(\sigma_i|\,\overline\sigma, \rho) = V_\lambda (\sigma_i) + \frac{\varepsilon}{2}\sigma_i^2 \rho({\bf r}_i,t) + (c-\varepsilon)\sigma_i \, \overline{\sigma}({\bf r}_i,t) \ ,
\end{equation}
with $c=(2\pi/L^2)\partial_{\varphi}U$. We have introduced the hydrodynamic fields of local density and magnetization, respectively denoted by $\rho$ and $\overline\sigma$, and defined as
\begin{equation}\label{eqn:fields}
    \rho(\mathbf{r},t) = \sum_{i=1}^N\delta(\mathbf{r}-\mathbf{r}_i(t)) \ ,
    \quad
    \overline{\sigma}(\mathbf{r},t)=\sum_{i=1}^N\sigma_i(t)\delta(\mathbf{r}-\mathbf{r}_i(t)) \ .
\end{equation}
The equations Eqs.~\eqref{eqn:effective_dyn_r} and~\eqref{eqn:effective_dyn_sigma} constitute the microscopic dynamics that we now set to coarse-grain into hydrodynamic equations.

Using stochastic calculus~\cite{Dean1996, FODOR2018}, the dynamics of the empirical joint distribution
\begin{equation}
    \psi(\mathbf{r},\sigma,t) = \sum_{i=1}^N\delta(\mathbf{r}_i(t)-\mathbf{r})\delta(\sigma_i(t)-\sigma)
\end{equation}
can be straightforwardly deduced from the microscopic dynamics [Eqs.~\eqref{eqn:effective_dyn_r} and \eqref{eqn:effective_dyn_sigma}]:
\begin{equation}\label{eqn:hydro_f}
    \partial_t \psi = \mu_\sigma \partial_{\sigma}( \psi \partial_{\sigma}\Phi ) + \mu_{\sigma}T\partial_{\sigma}^2\psi + \mu_\mathbf{r} T \nabla^2 \psi \ ,
\end{equation}
where we have neglected the hydrodynamic noises. The hydrodynamic fields [Eq.~\eqref{eqn:fields}] can be expressed in terms of $\psi$ as
\begin{equation}
    \rho(\mathbf{r},t) = \int d{\sigma} \psi(\mathbf{r},\sigma,t) \ ,
    \quad
    \overline{\sigma}(\mathbf{r},t) =\int d{\sigma}\sigma \psi(\mathbf{r},\sigma,t) \ ,
\end{equation}
so that, integrating the hydrodynamics [Eq.~\eqref{eqn:hydro_f}] over $\sigma$, we deduce
\begin{equation}\label{eq:hydro_bis}
    \partial_t\rho = \mu_{\bf r} T\nabla^2\rho \ ,
    \quad
    \partial_t\overline{\sigma} = \mu_{\bf r} T \nabla^2 \overline{\sigma} - \mu_\sigma \int d\sigma \psi \partial_{\sigma}\Phi \ .
\end{equation}
The density field $\rho$ relaxes towards the homogeneous profile $\rho=\rho_0$. The last term in Eq.~\eqref{eq:hydro_bis} has to be determined explicitly to close the hydrodynamics. To this end, inspired by~\cite{Archer2009}, we consider a local steady-state ansatz $\psi \simeq \psi_{\rm ls}$ defined as
\begin{equation}\label{eq:ansatz}
    \psi_{\rm ls} = \frac{1}{Z(\overline{\sigma})} e^{-\beta[\Phi(\sigma|\,\overline{\sigma})-\chi(\overline{\sigma})\sigma]} \ ,
    \quad
    Z(\overline{\sigma}) = \int d{\sigma}e^{-\beta[\Phi(\sigma|\,\overline{\sigma}) - \chi(\overline{\sigma})\sigma]} \ ,
\end{equation}
where we have omitted the dependence on density $\rho=\rho_0$. The bias $\chi$ is defined by the self-consistent condition
\begin{equation}\label{eqn:sigbar}
    \overline{\sigma} = \frac{1}{Z(\overline{\sigma})} \int d{\sigma}\sigma e^{-\beta[ \Phi(\sigma|\,\overline{\sigma}) - \chi(\overline{\sigma})\sigma]} \ .
\end{equation}
The dynamics of $\overline\sigma$ directly follows by substituting the ansatz [Eq.~\eqref{eq:ansatz}] into Eq.~\eqref{eq:hydro_bis}:
\begin{equation}\label{eqn:hydro}
    \partial_t\overline{\sigma} = \mu_{\bf r} T\nabla^2\overline{\sigma} -\mu_\sigma\chi(\overline{\sigma}) \ .
\end{equation}
Therefore, we have obtained a closed hydrodynamics [Eqs.~\eqref{eqn:sigbar} and~\eqref{eqn:hydro}] for the magnetization field $\overline\sigma$. Interestingly, our coarse-graining amounts to an equilibrium mapping, since Eq.~\eqref{eqn:hydro} can actually be written as a Model-A dynamics~\cite{Bray1994}
\begin{equation}
    \partial_t\overline \sigma = - \mu_\sigma \frac{\delta\cal F}{\delta\overline\sigma} \ ,
\end{equation}
in terms of the free-energy functional
\begin{equation}\label{eqn:Model_A_FE}
    \mathcal{F}[\overline{\sigma}] = \int d \mathbf{r} \bigg[ \frac{\mu_\mathbf{r} T}{2\mu_\sigma}(\nabla\overline{\sigma})^2+f(\overline{\sigma})\bigg] \ ,
    \quad
    f(\overline{\sigma}) = \int^{\overline{\sigma}}\chi(\sigma) d \sigma \ .
\end{equation}
The gradient term in $\cal F$ penalizes the formation of interfaces, and the free-energy density $f$ determines the thermodynamic stability of a given $\overline\sigma$: the steady state then corresponds to the homogeneous profile given by the minimum of $f$. In the following Section, we analyze the corresponding transitions between various homogeneous configurations.


\begin{figure}
    \centering
    \includegraphics[width=\linewidth]{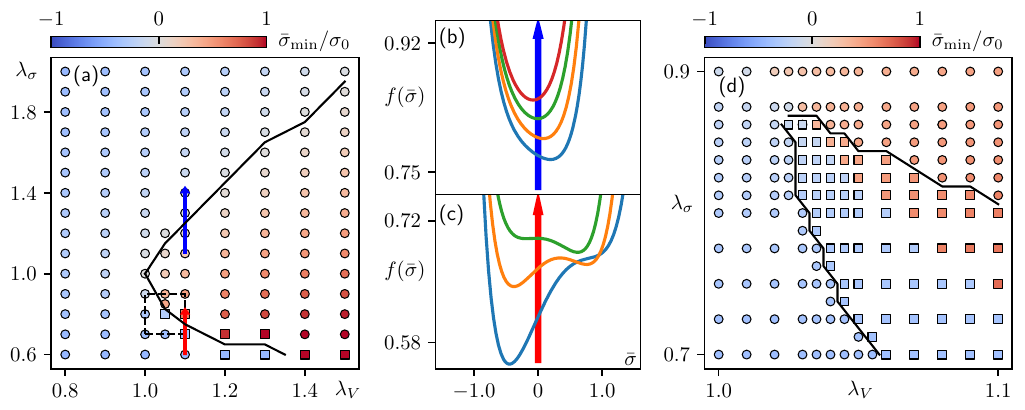}    
    \caption{(a)~Phase diagram of the stable magnetization $\overline\sigma_{\rm min}$ [Eq.~\eqref{eq:magn}] in terms of the landscape parameters $(\lambda_\sigma,\lambda_V)$. The black solid line corresponds to $\overline\sigma_{\rm min}=0$. The dashed solid line refers to the parameter regime shown in panel~(d).
    (b, c)~Free-energy density $f$ [Eq.~\eqref{eqn:Model_A_FE}] of the magnetization $\overline\sigma$ for various values of $\lambda_\sigma$ at fixed $\lambda_V=1.1$. The blue and red arrows refer to the parameter values in panel~(a).
    (d)~Phase diagram close to critical point $(\lambda_V,\lambda_\sigma)_c\approx (1.025,0.869)$. The solid black lines separate regimes where $f$ features either one minimum (circles) or two minima (squares).
    Parameters: $\rho=1.4$, $\varepsilon=1$, $C=0$, $\sigma_0=1$, $\sigma_{\neq}=0.4$ ($\alpha=0.12$).
    }
    \label{Fig:6}
\end{figure}

\subsection{Phase diagrams and transitions}

Based on the free-energy density $f(\sigma)$ [Eq.~\eqref{eqn:Model_A_FE}], our aim is to obtain a phase diagram that reports the steady-state magnetization $\overline\sigma$ as a function of ${\bm\lambda}=(\lambda_\sigma,\lambda_V)$. To this end, we determine the explicit shape $f(\sigma)$ using a numerical scheme, as detailed in~\ref{app:A}, from which we deduce the stable magnetization as
\begin{equation}\label{eq:magn}
    \overline\sigma_{\rm min} = \underset{\sigma}{\rm argmin} \,f(\sigma) .
\end{equation}
The cases of positive and negative magnetization distinguish different regimes of species survival: only $A$ (resp.~$B$) survives for $\overline\sigma_{\rm min}>0$ (resp.~$\overline\sigma_{\rm min}<0$). Remarkably, the phase boundary between these regimes features a qualitatively similar behavior for the hydrodynamic theory [$\overline\sigma_{\min}=0$ in Fig.~\ref{Fig:6}(a)] and its microscopic counterpart [$\langle\phi_A\rangle_s=1/2$ in Fig.~\ref{Fig:4}(b)]; note that the boundary crosses ${\bm\lambda}=(1,1)$ in both cases, as expected. This qualitative agreement clearly supports the validity of our coarse-graining from the microscopic dynamics to the equilibrium field theory.

At fixed $\lambda_V$, we observe a re-entrant behavior with two crossings of the phase boundary: increasing $\lambda_\sigma$ first leads from $\overline\sigma_{\rm min}<0$ to $\overline\sigma_{\rm min}>0$, and then to $\overline\sigma_{\rm min}<0$ [Fig.~\ref{Fig:6}(a)], which mirrors the microscopic case [Fig.~\ref{Fig:4}(a)]. Interestingly, the first crossing is a first-order transition with exchange of local free-energy minima [Fig.~\ref{Fig:6}(b)], whereas the second one simply amounts to a shift of the global minimum [Fig.~\ref{Fig:6}(c)]: the former is associated with an abrupt change of magnetization, while the latter corresponds to a smooth cross-over between positive and negative magnetizations. Consequently, there exists a critical point ${\bm\lambda}_c$ where the shape of the free energy density $f(\sigma)$ becomes locally concave. In practice, we locate ${\bm\lambda}_c$ as the meeting point between the two branches in the plane ${\bm\lambda}=(\lambda_\sigma,\lambda_V)$ distinguishing the regimes where $f$ has either one or two minima [Fig.~\ref{Fig:6}(d)].

\begin{figure}
    \centering
    \includegraphics[width=\linewidth]{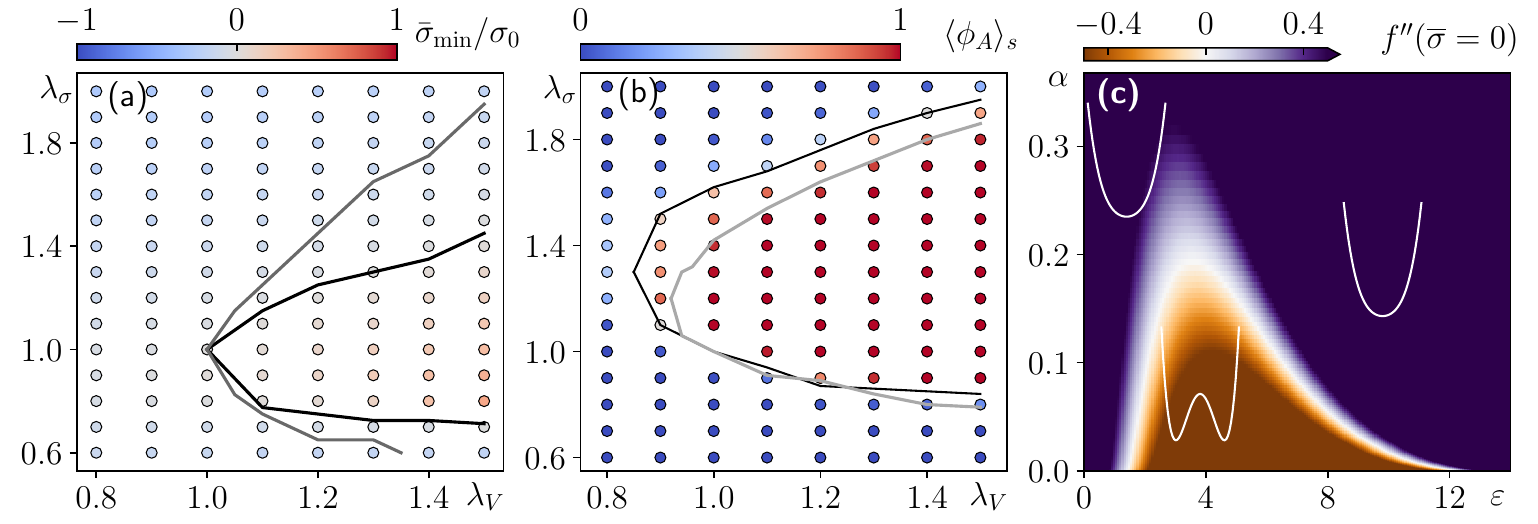}    
    \caption{(a)~Phase diagram of the stable magnetization $\overline\sigma_{\rm min}$ for $\alpha=0.3$, other parameters in as Fig.~\ref{Fig:6}(a). The black solid line corresponds to $\overline\sigma_{\rm min}=0$ for $\alpha=0.3$, and the gray solid line for $\alpha=0.12$ [Fig.~\ref{Fig:6}(a)].
    (b)~Phase diagram of the steady-state population $\langle\phi_A\rangle_s$ from particle-based dynamics for $\alpha=0.23$, other parameters as in Fig.~\ref{Fig:4}(b). The black solid line corresponds to $\langle\phi_A\rangle_s=0.5$ for $\alpha=0.23$, and the gray solid line for $\alpha=0.12$ [Fig.~\ref{Fig:4}(b)].
    (c)~Curvature of the free-energy density $f''=(d^2f/d\overline\sigma^2)(\overline{\sigma}=0)$ [Eq.~\eqref{eqn:lincoeff}] as a function of the synchronization strength $\varepsilon$ and the metastability parameter $\alpha$. The case $f''=0$ distinguishes regimes with either one minimum or two minima in $f$. Parameters: $\lambda_\sigma=1$, $\lambda_V=1$, $c=0$, $\rho_0=1.4$, $\beta=1$, $\sigma_0=1$.
    }
    \label{Fig:7}
\end{figure}

Remarkably, although the free-energy density $f$ captures the stability of enantiomers, which mirrors the stability of the corresponding minima in the microscopic landscape $V_\lambda$ [Eq.~\eqref{eq:v_lam}], $f$ does not feature any metastable state close to $\sigma=0$ in contrast with $V_\lambda$. However, the relative depth $\alpha$ [Eq.~\eqref{eq:alpha}], which determines the metastability of the point particles, affects the shape of $f$ and therefore controls the survival of species. In practice, we find that varying $\alpha$ changes the phase boundary at the hydrodynamic level [Fig.~\ref{Fig:7}(a)] in a qualitatively similar way as in the microscopics [Fig.~\ref{Fig:7}(b)]. Importantly, the critical point ${\bm\lambda}_c$ is pushed further away from ${\bm\lambda}=(1,1)$ as $\alpha$ increases. To rationalize these observations, our aim is to obtain an explicit criterion for $f$ to become locally concave. In practice, we focus on the symmetric point ${\bm\lambda}=(1,1)$ where the free energy always admits a local minimum at $\sigma=0$; indeed, $\chi(\overline\sigma=0)=0$ [Eq.~\eqref{eqn:sigbar}], so that $(df/d\overline\sigma)_{\overline\sigma=0}=0$ [Eq.~\eqref{eqn:Model_A_FE}]. The existence of a critical point then amounts to
\begin{equation}\label{eq:f''}
	\frac{d^2 f}{d\overline\sigma^2}\bigg|_{\overline\sigma=0}=0 \quad {\rm at} \quad {\bm\lambda}=(1,1) \ .
\end{equation}
We start by expanding the integrand in the self-consistent relation [Eq.~\eqref{eqn:sigbar}]:
\begin{equation}
	e^{- \beta [ \Phi(\sigma|\overline\sigma) - \chi(\overline\sigma) \sigma ]} = e^{- \beta [ \Phi(\sigma|0) - \chi(0) \sigma ]} \bigg[ 1 - \beta \overline\sigma  \bigg( \frac{d\Phi}{d\overline\sigma} - \sigma \, \frac{d\chi}{d\overline\sigma} \bigg)_{\overline\sigma=0} + {\cal O}(\overline\sigma^2) \bigg] \ ,
\end{equation}
Using the explicit expression for the effective landscape $\Phi$ [Eq.~\eqref{eq:land}], and considering the symmetric case ${\bm\lambda}=(1,1)$, we deduce
\begin{equation}\label{eq:bolt}
	e^{- \beta [ \Phi(\sigma|\overline\sigma) - \chi(\overline\sigma) \sigma ]} = e^{- \beta [ V(\sigma) + \varepsilon \rho_0 \sigma^2 / 2 ]} \bigg[ 1 -  \beta \overline\sigma \sigma \bigg( c-\varepsilon - \frac{d^2 f}{d\overline\sigma^2}\bigg|_{\overline\sigma=0} \bigg) + {\cal O}(\overline\sigma^2) \bigg] \quad {\rm at} \quad {\bm\lambda}=(1,1) \ ,
\end{equation}
where we have used that $\chi(\overline\sigma=0)=0$ [Eq.~\eqref{eqn:sigbar}], $(d\chi/d\overline\sigma)_{\overline\sigma=0}=f''$ [Eqs.~\eqref{eqn:Model_A_FE} and~\eqref{eq:f''}], and $V_\lambda=V$ [Eq.~\eqref{eq:v_lam}] for ${\bm\lambda}=(1,1)$. Finally, substituting the expansion of the integrand [Eq.~\eqref{eq:bolt}] into the self-consistent relation [Eq.~\eqref{eqn:sigbar}], we deduce
\begin{equation}\label{eqn:lincoeff}
	\frac{d^2 f}{d\overline\sigma^2}\bigg|_{\overline\sigma=0} = c - \varepsilon + \frac{\int d{\sigma}e^{-\beta[V(\sigma) + \varepsilon \rho_0 \sigma^2 / 2]}}{\beta\int d{\sigma}\sigma^2e^{-\beta[V(\sigma) + \varepsilon \rho_0 \sigma^2 / 2]}} \quad {\rm at} \quad {\bm\lambda}=(1,1) \ .
\end{equation}
Substituting Eq.~\eqref{eqn:lincoeff} into the condition of Eq.~\eqref{eq:f''} yields a relation for criticality to emerge at ${\bm\lambda}=(1,1)$. In practice, increasing $\alpha$ at fixed $\varepsilon$ always changes the shape of $f$ from one to two minima [Fig.~\ref{Fig:7}(c)]: this is consistent with our observation that the critical point moves away from ${\bm\lambda}=(1,1)$ as $\alpha$ increases [Fig.~\ref{Fig:7}(a)].


\section{Discussion}\label{sec:conclusions}

We introduce a particle-based model in which the interconversion between species is realized through particle deformation. We demonstrate that a transient phase separation controls the relaxation towards a homogeneous configuration where only a single species survives. We rationalize the competition between species in terms of some specific kinetic factors of relaxation, and recapitulate the corresponding phase diagram with a hydrodynamic mapping to equilibrium given by Model-A dynamics for the magnetization field. The effective free energy is obtained through a self-consistent relation based on purely static arguments, and successfully reproduces the steady-state of microscopic dynamics. It would be interesting to study further whether our hydrodynamics also captures the distinct relaxation stages observed in particle-based simulations.

Although the phase separation is only transient, it could be sustained by a periodic driving. It would be interesting to explore whether driving the parameters of the asymmetric landscape can interrupt the relaxation and thus maintain phase separation. Specifically, our phase diagram can already serve to delineate the boundaries of some cyclic protocols in parameter space. In fact, we expect that periodically driving the system between regimes of survival of $(A,B)$ would lead domains of $(A,B)$ to switch between growth and shrinking periods without ever fully relaxing. In this context, enforcing a driving period shorter than the internal relaxation is a route to sustaining phase separation. Non-equilibrium periodic protocols could be built on top of the equilibrium dynamics of responsive colloids driven by external fields~\cite{Moncho24jcp}. In practice, such periodic protocols would mimic the scenario recently observed in intracellular phase separation~\cite{yan2022condensate, Charras2022}, where condensates with finite sizes maintain themselves only due to periodic changes of their environment.

An interesting perspective would then be to optimize the periodic driving of the landscape. To this end, one could build on recent works addressing such an optimization for microscopic heat engines~\cite{Frim2022, Frim2022b}. In the context of species interconversion, the cost function could be either the work produced or the heat spent by the driving, as defined through stochastic thermodynamics~\cite{Seifert_2012, Cates2022}. Interestingly, some insights from response theory provide a systematic approach to optimizing the control of stochastic systems though external driving~\cite{Sivak2012, Davis2024}. Here, such an optimization could be deployed to either the dynamics of deformable particles, or its hydrodynamic counterpart. Yet, our hydrodynamic equilibrium mapping discards important nonequilibrium contributions, which makes it inconsistent with the microscopics at the thermodynamic level. Proposing a thermodynamically consistent coarse-graining, which accounts for all sources of dissipation, should be the first step towards a proper optimization.

\section*{Acknowledgements}

A. M. has received funding from the European Union’s Horizon Europe program under the Marie Sk\l{}odowska-Curie Action Grant no.~101056825 (NewGenActive) and from the project MOCA funded by MUR PRIN2022 grant No. 2022HNW5YL. Work funded in part by the Luxembourg National Research Fund (FNR), grant reference 14389168.


\appendix
\section{Numerical evaluation of free-energy density}\label{app:A}

In this Appendix, we describe here how to obtain the free-energy density $f(\sigma)=\int^\sigma d\sigma'\chi(\sigma')$ [Eq.~\eqref{eqn:Model_A_FE}]. To this end, we numerically solve through a standard gradient-descent procedure [Algorithm~\ref{alg:GD}] the self-consistent relation for the magnetization $\overline\sigma$ [Eq.~\eqref{eqn:sigbar}] that defines the bias function $\chi(\overline\sigma)$. This procedure minimizes the cost function $\mathcal{L}$ quantifying the distance between the target magnetization $\overline\sigma$ and its self-consistent value $\overline\sigma_{\rm self}$. In practice, the parity of the effective landscape $\Phi(\sigma|\overline\sigma=0)=\Phi(-\sigma|\overline\sigma=0)$ [Eq.~\eqref{eq:land}] ensures $\chi(\overline\sigma=0)=0$ at the symmetric point ${\bm\lambda}=(1,1)$, where $V_\lambda=V$. We therefore use $\chi_0=0$ as the starting value of Algorithm~\ref{alg:GD} at small $\overline\sigma$, and use the converged solution $\chi(\overline\sigma)$ as the new starting value for $\chi(\overline\sigma +  d \overline\sigma)$.

\begin{algorithm}
\caption{Gradient-descent algorithm}\label{alg:GD}
\KwData{$\overline{\sigma}$}
\KwResult{$\chi(\overline{\sigma})$}
$\overline{\sigma} \gets \overline{\sigma}$

$\chi \gets \chi_0$

$\mathcal{L}\gets1$

\While{$\mathcal{L} \ge \mathcal{L}_{c}$}{
    $\chi \gets \chi - \partial_\chi\mathcal{L}$
    
    $Z \gets \int d{\sigma}e^{-\beta [\Phi(\sigma|\,\overline{\sigma})-\chi\sigma]}$
    
    $\overline\sigma_{\rm self} \gets Z^{-1}\int d{\sigma}\sigma e^{-\beta[\Phi(\sigma|\,\overline{\sigma})-\chi\sigma]}$
    
    ${\cal L}\gets (\overline\sigma_{\rm self}-\overline{\sigma})^2$
}
$\chi(\overline \sigma) \gets \chi$
\end{algorithm}


\section*{References}

\bibliographystyle{iopart-num}
\bibliography{draft}

\providecommand{\newblock}{}
\begin{thebibliography}{10}
\expandafter\ifx\csname url\endcsname\relax
  \def\url#1{{\tt #1}}\fi
\expandafter\ifx\csname urlprefix\endcsname\relax\def\urlprefix{URL }\fi
\providecommand{\eprint}[2][]{\url{#2}}

\bibitem{Hohenberg1977}
Hohenberg P~C and Halperin B~I 1977 {\em Rev. Mod. Phys.\/} {\bf 49} 435--479

\bibitem{Bray1994}
Bray A~J 1994 {\em Adv. Phys.\/} {\bf 43} 357--459

\bibitem{Bocquet1996}
Bocquet J, Brebec G and Limoge Y 1996 Chapter 7 - {Diffusion} in metals and
  alloys {\em Physical Metallurgy (Fourth Edition)\/} ed Cahn R~W and Haasen P
  (Oxford: North-Holland) pp 535--668 fourth edition ed ISBN 978-0-444-89875-3

\bibitem{Dehosson2001}
{De Hosson} J~T~M and Kooi B~J 2001 Chapter 1 - {Microstructure} and properties
  of interfaces between dissimilar materials {\em Handbook of Surfaces and
  Interfaces of Materials\/} ed Nalwa H~S (Burlington: Academic Press) pp
  1--113 ISBN 978-0-12-513910-6

\bibitem{Bernu87pra}
Bernu B, Hansen J~P, Hiwatari Y and Pastore G 1987 {\em Phys. Rev. A\/} {\bf
  36} 4891--4903

\bibitem{Grigera01pre}
Grigera T~S and Parisi G 2001 {\em Phys. Rev. E\/} {\bf 63} 045102

\bibitem{Schelling1971}
Schelling T~C 1971 {\em J. Math. Sociol.\/} {\bf 1} 143--186

\bibitem{Hyman2014}
Hyman A~A, Weber C~A and J{\"{u}}licher F 2014 {\em Annu. Rev. Cell Dev.
  Biol.\/} {\bf 30} 39--58

\bibitem{Shin2017}
Shin Y and Brangwynne C~P 2017 {\em Science\/} {\bf 357} 6357

\bibitem{Weber19rpp}
Weber C~A, Zwicker D, Jülicher F and Lee C~F 2019 {\em Rep. Prog. Phys.\/}
  {\bf 82} 064601

\bibitem{Shrinivas2021}
Shrinivas K and Brenner M~P 2021 {\em Proc. Natl. Acad. Sci. U.S.A.\/} {\bf
  118} 1--8

\bibitem{Thewes22prl}
Thewes F~C, Kr\"uger M and Sollich P 2023 {\em Phys. Rev. Lett.\/} {\bf 131}
  058401

\bibitem{Grosberg15pre}
Grosberg A~Y and Joanny J~F 2015 {\em Phys. Rev. E\/} {\bf 92} 032118

\bibitem{Weber16prl}
Weber S~N, Weber C~A and Frey E 2016 {\em Phys. Rev. Lett.\/} {\bf 116} 058301

\bibitem{Ilker20prr}
Ilker E and Joanny J~F~m~c 2020 {\em Phys. Rev. Res.\/} {\bf 2} 023200

\bibitem{Alston2022}
Alston H, Parry A~O, Voituriez R and Bertrand T 2022 {\em Phys. Rev. E\/} {\bf
  106} 034603

\bibitem{berthin2024}
Berthin R, Fries J, Jardat M, Dahirel V and Illien P 2024   arXiv:2406.14256

\bibitem{Togashi2019}
Togashi Y 2019 {\em J. Phys. Chem. B\/} {\bf 123} 1481--1490

\bibitem{Zhang23prl}
Zhang Y and Fodor E 2023 {\em Phys. Rev. Lett.\/} {\bf 131} 238302

\bibitem{manacorda2023}
Manacorda A and Étienne Fodor 2023   arXiv:2310.14370

\bibitem{pineros2024}
Piñeros W~D and Étienne Fodor 2024   arXiv:2403.16961

\bibitem{Dzubiella2024}
Göth N and Dzubiella J 2024   arXiv:2408.11560

\bibitem{Ising1925}
Ising E 1925 {\em Z. Physik\/} {\bf 31} 253–258

\bibitem{Glauber1963}
Glauber R~J 1963 {\em J. Math. Phys.\/} {\bf 4} 294--307

\bibitem{Latinwo2016}
Latinwo F, Stillinger F~H and Debenedetti P~G 2016 {\em J. Chem. Phys.\/} {\bf
  145} 154503

\bibitem{Stillinger2023}
Piaggi P~M, Car R, Stillinger F~H and Debenedetti P~G 2023 {\em J. Chem.
  Phys.\/} {\bf 159} 114502

\bibitem{yan2022condensate}
Yan V~T, Narayanan A, Wiegand T, J{\"u}licher F and Grill S~W 2022 {\em
  Nature\/} {\bf 609} 597--604

\bibitem{Charras2022}
Charras G and Lenz M 2022 {\em Nature\/} {\bf 609} 469--470

\bibitem{Odor2008}
\'Odor G 2004 {\em Rev. Mod. Phys.\/} {\bf 76} 663--724

\bibitem{Ziff1986}
Ziff R~M, Gulari E and Barshad Y 1986 {\em Phys. Rev. Lett.\/} {\bf 56}
  2553--2556

\bibitem{Bramson88prl}
Bramson M and Lebowitz J~L 1988 {\em Phys. Rev. Lett.\/} {\bf 61} 2397--2400

\bibitem{Zhuo1993}
Zhuo J, Redner S and Park H 1993 {\em J. Phys. A\/} {\bf 26} 4197

\bibitem{Brown97pre}
Brown K~S, Bassler K~E and Browne D~A 1997 {\em Phys. Rev. E\/} {\bf 56}
  3953--3958

\bibitem{cahn1958free}
Cahn J~W and Hilliard J~E 1958 {\em J. Chem. Phys.\/} {\bf 28} 258--267

\bibitem{Li20jstatmech}
Li Y~I and Cates M~E 2020 {\em J. Stat. Mech.: Theory Exp.\/} {\bf 2020} 053206

\bibitem{Anisimov2021}
Shumovskyi N~A, Longo T~J, Buldyrev S~V and Anisimov M~A 2021 {\em Phys. Rev.
  E\/} {\bf 103} L060101

\bibitem{Longo23pnas}
Longo T~J, Shumovskyi N~A, Uralcan B, Buldyrev S~V, Anisimov M~A and
  Debenedetti P~G 2023 {\em Proc. Natl. Acad. Sci. U.S.A.\/} {\bf 120}
  e2215012120

\bibitem{Milster23pre}
Milster S, Darwish A, G\"oth N and Dzubiella J 2023 {\em Phys. Rev. E\/} {\bf
  108}(4) L042601

\bibitem{Brito18prx}
Brito C, Lerner E and Wyart M 2018 {\em Phys. Rev. X\/} {\bf 8} 031050

\bibitem{Baul21jpcm}
Baul U and Dzubiella J 2021 {\em Journal of Physics: Condensed Matter\/} {\bf
  33} 174002

\bibitem{Gardiner2009}
Gardiner C~W 2009 {\em {Stochastic Methods: A Handbook for the Natural and
  Social Sciences}\/} (Springer)

\bibitem{Fodor22arcmp}
Fodor E, Jack R~L and Cates M~E 2022 {\em Annual Review of Condensed Matter
  Physics\/} {\bf 13} 215--238

\bibitem{banerjee2024}
Banerjee T, Desaleux T, Ranft J and Étienne Fodor 2024   arXiv:2407.19955

\bibitem{Dean1996}
Dean D~S 1996 {\em J. Phys. A\/} {\bf 29} 24

\bibitem{FODOR2018}
Fodor E and Cristina Marchetti M 2018 {\em Physica A\/} {\bf 504} 106--120

\bibitem{Archer2009}
Archer A~J 2009 {\em J. Chem. Phys.\/} {\bf 130} 014509

\bibitem{Moncho24jcp}
Moncho-Jord\'a A, Groh S and Dzubiella J 2024 {\em The Journal of Chemical
  Physics\/} {\bf 160} 024904

\bibitem{Frim2022}
Frim A~G and DeWeese M~R 2022 {\em Phys. Rev. Lett.\/} {\bf 128} 230601

\bibitem{Frim2022b}
Frim A~G and DeWeese M~R 2022 {\em Phys. Rev. E\/} {\bf 105} L052103

\bibitem{Seifert_2012}
Seifert U 2012 {\em Rep. Prog. Phys.\/} {\bf 75} 126001

\bibitem{Cates2022}
Fodor {\'E}, Jack R~L and Cates M~E 2022 {\em Annu. Rev. Condens. Matter
  Phys.\/} {\bf 13} 215--238

\bibitem{Sivak2012}
Sivak D~A and Crooks G~E 2012 {\em Phys. Rev. Lett.\/} {\bf 108} 190602

\bibitem{Davis2024}
Davis L~K, Proesmans K and Fodor E 2024 {\em Phys. Rev. X\/} {\bf 14} 011012

\bibitem{data}
Manacorda A 2025 Species interconversion of deformable particles yields
  transient phase separation \urlprefix\url{osf.io/dev23}

\end{thebibliography}

\end{document}